# Dynamical CP violation at finite temperature

WANG Dian-Fu（王殿夫）

Department of Physics, Dalian Maritime University,
Dalian 116026; E-mail: wangdfu@dlmu.edu.cn

SUN Xiao-Yu（孙小雨）

School of Physics and Electronic Technology,
Liaoning Normal University, Dalian 116029

LIANG Chao（梁超）

Department of Physics, Dalian Maritime University, Dalian 116026

## Abstract

Based on the generalized Yang-Mills model, CP violation behavior at finite temperature is investigated, and it is shown that dynamical CP violation of the generalized Yang-Mills model at zero temperature can be restored at finite temperature.

# 1  Introduction

As is known in 1956, Lee and Yang [1] proposed that parity symmetry is no conservation in weak interactions. In 1957, Ramsey [2] and Jackson [3] further pointed out that T invariance was also an assumption and needed to be checked by experiments. And then in 1964, the phenomenon of CP violation was discovered by Val Fitch, James Cronin and collaborators [4] in $K^0$ decays. Following these works, the study of CP symmetry and its breaking has played a central role in particle physics.

In gauge theories CP violation has two aspects to it: The one is the CKM model [5,6]. The other is the strong CP violation. The first is a theoretical problem that must be solved in any realistic model of weak CP violation where the source of CP violation arises in the strong interaction sector of the theory from the term $\theta(\alpha_s/8\pi)G\tilde{G}$, which is of topological origin. It is well known that the CKM model does not provide a solution to the strong CP violation. And the CP violation in the CKM model is not sufficient to generate the desired amount of baryon asymmetry in cosmology. In addition to these there are other avenues which may reveal the existence of new source of CP violation that exists in the Standard Model. The largest values of the electric dipole moments in the framework of the Standard Model are very small. In the past decades a significant body of works on CP violation beyond the standard CKM model has appeared. It encompasses left-right symmetric models [7], spontaneous CP violation models [8-11], Timeon model [12] and so on.

There are indications that one needs a new source of CP violation above and beyond what is present in the Standard Model. Further, new sources of CP violation beyond the standard model could also show up in particles production at the LHC, and in the new generation of experiments underway on neutrino physics.

In Re.[13], we have constructed a maximally generalized Yang-Mills model. By considering the combination of the maximally generalized Yang-Mills model and the NJL mechanism [14], we have proposed a new mechanism of CP violation in Re.[15]. In the present paper, we will investigate the behavior of our previous CP violation model [15] at finite temperature. We show that the dynamical CP violation at zero temperature can be restored at finite temperature.

# 2  Dynamical CP violation of the generalized Yang-Mills model

In this section, we will briefly review the dynamical CP violation of the generalized Yang-Mills model which we have proposed in Re.[15]. By using the Maximally generalized Yang-Mills model [13] with $\Phi = S + i\gamma_5 P$, the Lagrangian of the generalized Yang-Mills model is given as

$$L = -\bar{\Psi}\gamma_\mu(\partial_\mu - iV_\mu)\Psi - \bar{\Psi}(Q_S S + iQ_P\gamma_5 P)\Psi - \frac{1}{2g^2}\tilde{Tr}(\partial_\mu V_\nu - \partial_\nu V_\mu - i[V_\mu, V_\nu])^2$$

$$-\frac{1}{4g^2}\tilde{Tr}\{Tr[\gamma_\mu\partial_\mu(S+i\gamma_5 P)-i\{\gamma_\mu V_\mu,(S+i\gamma_5 P)\}]^2\}. \tag{1}$$

This Lagrangian is invariant under CP and T transformations. For convenience, we neglect the interaction terms between the scalar field and the pseudoscalar field. Then the Lagrangian density (1) changes to be

$$L = -\bar{\Psi}\gamma_\mu(\partial_\mu - iV_\mu)\Psi - \bar{\Psi}(Q_S S + iQ_P\gamma_5 P)\Psi - \frac{1}{2g^2}\tilde{Tr}(\partial_\mu V_\nu - \partial_\nu V_\mu - i[V_\mu, V_\nu])^2$$

$$-\frac{1}{g^2}\tilde{Tr}\left[(\partial_\mu S - i\{V_\mu, S\})^2 + (\partial_\mu P - i[V_\mu, P])^2\right]. \tag{2}$$

Following Eq.(2), one can find that this model does not include the Higgs potential term $V$ as usual. Despite this, by using the NJL mechanism the symmetry breaking of this model can be realized dynamically.

From the Lagrangian (2), we can obtain the equation of motion as

$$\gamma_\mu(\partial_\mu - igV_\mu^a T^a)\Psi + (G_S S^c T^c + iG_P\gamma_5 P^b T^b)\Psi = 0, \tag{3}$$

$$(\partial_\mu^2 - g^2 d_S V_\mu^a V_\mu^a)S^c - G_S\bar{\Psi}T^c\Psi = 0, \tag{4}$$

$$(\partial_\mu^2 - g^2 f_P V_\mu^a V_\mu^a)P^b - iG_P\bar{\Psi}\gamma_5 T^b\Psi = 0, \tag{5}$$

$$(\partial_\mu F_{\mu\nu}^a + gf^{abc}V_\mu^b F_{\mu\nu}^c) + [g^2 d_S(S^c)^2 + g^2 f_P(P^b)^2]V_\nu^a - ig\bar{\Psi}\gamma_\nu T^a\Psi = 0, \tag{6}$$

in which $G_S = gQ_S$, $G_P = gQ_P$, $F_{\mu\nu}^a = \partial_\mu V_\nu^a - \partial_\nu V_\mu^a + gf^{abc}V_\mu^b V_\nu^c$ and $d_S = d^{abc}d^{abc}$, $f_P = f^{abc}f^{abc}$ (in $d_S$ and $f_P$, where $a$ varies from 1 to $N_V$, $b$ varies from $N_V+1$ to $N_V+N_P$, and $c$ varies from $N_V+N_P+1$ to $N_V+N_P+N_S$). Multiplying the left and right-hand side of Eq.(6) by $V_\nu^a$, and then taking vacuum expectation value of it, to the lowest-order approximation in $\hbar$, we have[16-18]

$$f_V\langle V_\mu^a V_\mu^a\rangle = d_S\langle(S^c)^2\rangle + f_P\langle(P^b)^2\rangle, \tag{7}$$

with $f_V = f^{abc}f^{abc}$ ($a,b,c$ vary from 1 to $N_V$). Here one can choose the scalar bosons $S^{c_1}$ and $P^{b_1}$ to be associated with the unit generator $T^{c_1} = T^{b_1} = 1/\sqrt{2N_d}$ times the $N_d \times N_d$ unit matrix ($N_d$: the dimensions of the fundamental representation), and if there is no unit generator in the Lie group one can introduce a unit one to it. We denote the vacuum expectation of the scalar bosons

$$\langle S^c\rangle = \langle S^{c_1}\rangle \neq 0, \quad \langle P^b\rangle = \langle P^{b_1}\rangle \neq 0. \tag{8}$$

Seeing from the Lagrangian (2), we can conclude that the nonzero expectation value of the pseudoscalar field $P^b$ implies CP violation.

Let's perform a unitary transformation $\Psi \to e^{-i\frac{1}{2}\gamma_5 \alpha}\Psi$ under which $P^b$ is unchanged. Hence, in equation (3), by choosing

$$\tan\alpha = \frac{G_P \langle P^b \rangle T^b}{G_S \langle S^c \rangle T^c}, \tag{9}$$

we have

$$\bar{\Psi}(G_S \langle S^c \rangle T^c + iG_P \gamma_5 \langle P^b \rangle T^b)\Psi \to \bar{\Psi} M_\Psi \Psi, \tag{10}$$

in which the fermion mass

$$M_\Psi = \left[\frac{1}{2N_d}(G_S \langle S^{c_1} \rangle)^2 + \frac{1}{2N_d}(G_P \langle P^{b_1} \rangle)^2\right]^{\frac{1}{2}}. \tag{11}$$

Then after taking the vacuum expectation values of Eq.(4) and Eq.(5) to the lowest-order approximation in $\hbar$, we obtain the self-consistency equations as

$$M_S^2 \langle S^{c_1} \rangle = -G_S \sqrt{\frac{1}{2N_d}} \langle \bar{\Psi}\Psi \rangle = iG_S \sqrt{\frac{1}{2N_d}} Tr S_F(0), \tag{12}$$

$$M_P^2 \langle P^{b_1} \rangle = -iG_P \sqrt{\frac{1}{2N_d}} \langle \bar{\Psi}\gamma_5 \Psi \rangle = -G_P \sqrt{\frac{1}{2N_d}} Tr[\gamma_5 S_F(0)], \tag{13}$$

In which $M_S^2 = g^2 d_S \langle V_\mu^a V_\mu^a \rangle$, $M_P^2 = g^2 f_P \langle V_\mu^a V_\mu^a \rangle$.

From Eq.(12) and Eq.(13), we have

$$\langle S^{c_1} \rangle = -\frac{if_P G_S Tr S_F(0)}{d_S G_P Tr[\gamma_5 S_F(0)]} \langle P^{b_1} \rangle = -\frac{f_P G_S}{d_S G_P \tan\alpha} \langle P^{b_1} \rangle. \tag{14}$$

Substituting Eq.(14) into Eq.(11), we obtain

$$M_\Psi = \eta \langle P^{b_1} \rangle, \tag{15}$$

with $\eta = \frac{1}{d_S G_P \tan\alpha} \sqrt{\frac{f_P^2 G_S^4 + d_S^2 G_P^4 \tan^2\alpha}{2N_d}}$. Then substituting Eq.(7) and Eq.(14) into Eq.(13), we can rewrite the self-consistency equation (13) as

$$(g^2 f_P^3 G_S^2 + g^2 d_S f_P^2 G_P^2 \tan^2\alpha)\langle P^{b_1} \rangle^3 = -f_V d_S G_P^3 \tan^2\alpha \sqrt{\frac{1}{2N_d}} Tr[\gamma_5 S_F(0)]. \tag{16}$$

By using Eq.(14), Eq.(15) and Eq.(16), we can finally obtain the non-vanishing vacuum expectation values of the pseudoscalar field $P$ which is completed determined by the self-energy of the fermions. So the CP and T symmetry is broken down dynamically.

## 3 CP violation at finite temperature

In this section we will investigate the behavior of CP violation when we take temperature into account. Let us now assume that the system is at finite temperature $T = 1/\beta$. In this case we can use Matsubara formalism which consists in taking

$p_0 = (2n+1)\pi i/\beta$ (n integer for fermions) and changing $(1/2\pi)\int dp_0 \to \frac{i}{\beta}\sum_{n=-\infty}^{\infty}$ [19].

And at the same times we also change the system from Minkowski space to Euclidean space. Then the self-consistency equation (16) can be written as

$$[g^2 f_P^3 G_S^2 + g^2 d_S f_P^2 G_P^2 \tan^2 \alpha(\beta)]\langle P^{b_1}(\beta)\rangle^2$$

$$= -f_V d_S G_P^3 \tan^2 \alpha(\beta)\sqrt{\frac{1}{2N_d}}Tr\left[\gamma_5 S_F(0)_\beta\right]$$

$$= f_V d_S G_P^4 \tan^2 \alpha(\beta)\frac{4}{2N_d}\frac{1}{\beta}\sum_n \int \frac{d^3\bar{p}}{(2\pi)^3}\frac{-1}{p^2 + \eta^2(\beta)\langle P^{b_1}(\beta)\rangle^2}$$

$$= \frac{f_V d_S G_P^4}{2N_d}\tan^2 \alpha(\beta)(A+B), \tag{17}$$

with

$$A = \frac{1}{\pi^2}\int_0^\infty \frac{|\bar{p}|^2 d|\bar{p}|}{\sqrt{|\bar{p}|^2 + \eta^2(\beta)\langle P^{b_1}(\beta)\rangle^2}}, \tag{18}$$

$$B = -\frac{2}{\pi^2}\int_0^\infty \frac{|\bar{p}|^2 d|\bar{p}|}{\sqrt{|\bar{p}|^2 + \eta^2(\beta)\langle P^{b_1}(\beta)\rangle^2}\left\{\exp\left[\beta\sqrt{|\bar{p}|^2 + \eta^2(\beta)\langle P^{b_1}(\beta)\rangle^2}\right]+1\right\}}. \tag{19}$$

The integration in $A$ is divergent. By introducing an invariant momentum cut-off at $p^2 = \Lambda^2$, the integration will be finite as

$$A = \frac{1}{2\pi^2}\left[\Lambda^2 - \eta^2(\beta)\langle P^{b_1}(\beta)\rangle^2 \ln\left(\frac{\Lambda^2}{\eta^2(\beta)\langle P^{b_1}(\beta)\rangle^2}+1\right)\right]. \tag{20}$$

We are interested in the CP symmetry behavior at finite temperature and wish to find the critical temperatures $T_C$ at which the order parameter $<P^{b_1}(\beta_C)>$ tends to zero. So the integration in $B$ can be calculated in the approximation $<P^{b_1}(\beta_C)>=0$ as

$$B = -\frac{2}{\pi^2}\int_0^\infty \frac{|\bar{p}|d|\bar{p}|}{1+\exp \beta_C|\bar{p}|} = -\frac{1}{6\beta_C^2}. \tag{21}$$

And in the approximation $<P^{b_1}(\beta_C)>=0$, the integration in $A$ changes to be

$$A = \frac{\Lambda^2}{2\pi^2}. \qquad (22)$$

Substituting Eq.(21) and Eq.(22) into Eq.(17) (with the limit $<P^{b_1}(\beta_C)> \to 0$), we have

$$T_c = \frac{1}{\beta_c} = \frac{\sqrt{3}\Lambda}{\pi}. \qquad (23)$$

Thus, we have obtained the critical temperature $T_c$ at which the order parameter $<P^{b_1}(\beta_C)>$ tends to zero, which means that the dynamical breaking of the CP and T symmetry at zero temperature can be restored at finite temperature. This result is the same as those of Re.[17], Re.[20-22], which shows us that the critical temperature of the phase transformation in dynamical symmetry breaking model is only related to the momentum cut-off $\Lambda$.